 \UseRawInputEncoding
\documentclass[%
reprint,
superscriptaddress,
floatfix,
 amsmath,amssymb,amsfonts,
prl,
]{revtex4-2}

\usepackage{hyperref}
\usepackage{graphicx}
\usepackage{dcolumn}
\usepackage{bm, bbold}
\usepackage{comment, color}
\usepackage{natbib}
\usepackage{tabularx}

\usepackage[utf8]{inputenc}
\usepackage{lmodern}

\usepackage{hyperref}

\def\bea{\begin{eqnarray}}
\def\eea{\end{eqnarray}}
\def\beq{\begin{equation}}
\def\eeq{\end{equation}}
\def\f{\frac}

\def\be{\beta}

\def\h{\theta}

\def\a{\alpha}

\def\s{\sigma}

\def\la{\langle}
\def\ra{\rangle}
\def\nn{\nonumber}

\def\d{\delta}
\def\p{\partial}
\def\l{\lambda}

\def\w{\omega}
\def\g{\gamma}

\def\sv{ {\boldsymbol{\sigma}} }

\def\xv{ {\bf x}}

\def\uv{ {\bf u}}

\def\a{\alpha}
\def\d{\delta}
\def\p{\partial} 

\def\la{\langle}
\def\ra{\rangle}

\def\g{\gamma}

\def\hf{\frac{1}{2}}

\begin{document}

\title{Active stop and go motion: a strategy to improve spatial exploration and survival}

\author{Fernando Peruani}
\email{fernando.peruani@cyu.fr}
\address{Laboratoire de Physique Th{\'e}orique et Mod{\'e}lisation, UMR 8089, CY Cergy Paris Universit{\'e}, 95302 Cergy-Pontoise, France}

\author{Debasish Chaudhuri}
\email{debc@iopb.res.in}
\address{Institute of Physics, Sachivalaya Marg, Bhubaneswar 751005, India}
\address{Homi Bhabha National Institute, Training School Complex, Anushaktinagar, Mumbai 400094, India}

\date{\today}%

\begin{abstract}
We consider active Brownian particles that intermittently switch between active and inactive states. Such behavior is ubiquitous at all scales, from bacteria to animals and in artificial active systems. We derive exact expressions for key transport properties, including velocity autocorrelations and diffusion coefficients, and demonstrate that particle dispersion is highly sensitive to minute details, such as whether the memory of active orientation is retained across a stop event. Extending the model to include minimal metabolic dynamics, we show that the average survival time of the particle is maximized at (i)~an optimal stopping frequency and (ii)~a non-trivial optimal diffusivity, which itself depends on the stopping frequency.
 \end{abstract}

\maketitle

In most animal systems, active motion is not a continuous process but an intermittent one.  
Individuals move actively for a while before they pause and move again, restarting the cycle~\cite{kramer2001behavioral}. 
Intermittent active locomotion is often related to different ethological phases, such as moving in search of food or stopping to  rest~\cite{nathan2008movement}. 
Stop-and-go (S-\&-G) motion is also observed at the group level, with all group members moving and stopping in a synchronized fashion~\cite{tunstrom2013collective, trillmich2004coordination, ariel2014individual, gomez2022intermittent}. 
Intermittent motility patterns are even observed at a microscopic scale. The near-surface swimming bacteria, such as enterohemorrhagic {\it E. coli} perform  S-\&-G motion~\cite{perez2019bacteria}. The locomotion of gliding bacteria such as myxobacteria displays run and stop phases with ``stick-and-slip" statistics~\cite{gibiansky2013earthquake}. 
S-\&-G motion is also observed in active colloid systems subjected to an AC field~\cite{karani2019tuning, pradillo2019quincke}. 
Despite the experimental relevance of S-\&-G motion,  the large majority of active models have considered that particles are permanently active, i.e., their self-propulsion is always switched on~\cite{Romanczuk2012}.

We develop a unified theoretical framework for active particles, or active walkers, undergoing stop-and-go (S-\&-G) dynamics, showing that subtle microscopic details significantly impact large-scale transport and survival time -- the period over which internal energy of the particle stays positive -- by incorporating a minimal metabolic dynamics.
In particular, we analyze 4 representative models of experimental relevance within this framework. 
In model 1, the active heading direction of the particles fluctuates as the particle moves and is reset when the particle stops. 
This occurs, for instance, when bacteria in near-swimming perform transient adhesion events to a surface~\cite{perez2019bacteria} 
or in Quincke rollers driven by on-off-on electric field when the period is larger than the Maxwell-Wagner relaxation time~\cite{karani2019tuning, pradillo2019quincke}. 
In contrast, in model 2, after a stop, the particle continues moving in the direction it was moving prior to the stop.  
This behavior is observed in many insects and vertebrates that perform numerous short stops when moving~\cite{kramer2001behavioral, gomez2022intermittent}. These stops are supposed to serve to scan for the presence of predators~\cite{kramer2001behavioral, gomez2022intermittent}. 
Model 3 assumes that the heading direction fluctuates both in the moving and stop phases.
This is observed in cell migration, where fluctuations in cell polarization 
take place even when the cell is not moving~\cite{Schienbein1993, Vaidziulyte2022}.  
Finally, in model 4, the active moving direction is kept fixed while moving 
and evolves only during the stop phase. This behavior is often called  
 run-and-tumble motion and is characteristic of many bacterial systems~\cite{figueroa20203d, Wadhwa2022bacteria, kurzthaler2024characterization}.  

We incorporate an ecological dimension to the theoretical framework by adding minimal metabolic dynamics to active particles and assuming distributed food sources. This approach reveals an optimal stopping frequency and optimal diffusivity that maximize average survival time, offering fresh insights into the spatial dynamics of many organisms by unveiling a fundamental interplay between spatial exploration and survival in biological systems.

{\em Model.--} The state of the active walker, in any dimension, is described by $\{ \xv, \sv, q \}$, where the position $\xv$ is determined by 
\bea
\label{eq:xdot}
\dot \xv = v[q] \sv \equiv \uv(t) \, .
\eea
The  polarity 
$\sv$ specifies the active heading direction of the particle, and the behavioral state $q=\{S,M\}$ specifies the value of the active speed:  $v[q=S]=0$ and $v[q=M]=v_M$. 
The dynamics of $q$ follows: 
\bea
\label{eq:behavior}
M \overset{\be}{\underset{\g}\rightleftharpoons} S \, ,
\eea
where $\be$ and $\g$ are constant rates. 
The dynamics of $\sv$ is different for models 1 to 4. 
Note that,  in the final step,  we incorporate an internal energy variable $E$ into the state of the walker. 

\begin{table*}[t]
\begin{tabular}{|c|c|c|c|c|}
\hline
Models & Model 1 & Model 2 & Model 3 & Model 4\\ \hline
\hline
Rates: 1d & 
$\a_M=\a$, $\a_S=0$, reset
& 
$\a_M=\a$, $\a_S=0$
&  
$\a_M=\a_S=\a$
&  
$\a_M=0$, $\a_S=\a$\\ \hline
$D_{\rm eff}$ (1d) & $\f{\la v^2\ra}{2\a+\be}$ & $\f{\la v^2\ra}{2\a}$ & $\f{\la v \ra^2}{2\a} + \f{\la \d v^2\ra}{2\a+\be+\g}$ & $\la v^2\ra \left( \f{2\a+\g}{2\a \be}\right)$\\ \hline \hline
Rates: 2d & $D_M=D_r$, $D_S=0$, reset & $D_M=D_r$, $D_S=0$ & $D_M=D_S=D_r$  &  $D_M=0$, $D_S=D_r$ \\ \hline
$D_{\rm eff}$ (2d) & 
 $\hf\f{\la v^2\ra}{D_r+\be}$ & $\f{\la v^2\ra}{2 D_r}$ & $\f{\la v \ra^2}{2 D_r} + \hf \f{\la \d v^2\ra}{D_r+\be+\g}$ & $\la v^2\ra \left( \f{D_r+\g}{2 D_r \be}\right)$\\ \hline
Examples &  bacterial surface swimming~\cite{karani2019tuning, perez2019bacteria} & animal movement~\cite{kramer2001behavioral, gomez2022intermittent} & cell migration~\cite{Schienbein1993, Vaidziulyte2022} & bacterial run and tumble~\cite{figueroa20203d, Wadhwa2022bacteria, kurzthaler2024characterization} \\ \hline
\end{tabular}
\caption{Models and effective diffusion: Run-stop transitions are controlled by rates $\be,\, \g$~[Eq.(\ref{eq:behavior})].  Orientational persistence in the run state is controlled by rates $\a_M$ in 1D, $D_M$ in 2D, and in the stop state by $\a_S$ in 1D, $D_S$ in 2D.  The first and third rows indicate their specific choices. At reset, the stop state transitions to run with an equal probability of all possible orientations. The calculated effective diffusivities are listed in rows two and four. Examples corresponding to each model are cited in the fifth row. 
}
\label{table}
\end{table*}%

{\em One dimensional systems.--} In 1D, $\mathbf{x} \in \mathbb{R}$ and $\sv \in \{+1, -1\}$. 
The dynamics of the system can be described in terms of the probability distribution 
$P_q^{\sv}(x,t)$, 
of finding the particle at position $x$ at time $t$ in state $q$ and  polarity $\sv$. 
When the particle is in state $M$, the $\sv$-dynamics is given by $\sv \overset{\a_M}\longrightarrow -\sv$, 
while for state $S$ it becomes $\sv \overset{\a_S}\longrightarrow -\sv$, 
with $\a_{M,S}$ denoting transition rates between different orientations of heading direction.
The probability distributions $P_{\text{M}}^\pm(x,t)$ and $P_{\text{S}}^\pm(x,t)$ obey: 
\bea
\label{eq:main1d}
\p_t P_{\text{M}}^{+} &=& -v_M \p_x P_{\text{M}}^{+} -\a_M P_{\text{M}}^{+} + \a_M P_{\text{M}}^{-}  - \be P_{\text{M}}^{+} + \g P_{\text{S}}^{+}  \, ,\nn\\
\p_t P_{\text{M}}^{-} &=& \,\,\,\,\,v_M \p_x P_{\text{M}}^{-} - \a_M P_{\text{M}}^{-} + \a_M P_{\text{M}}^{+}  - \be P_{\text{M}}^{-} + \g P_{\text{S}}^{-} \, , \nn\\
\p_t P_{\text{S}}^{+} &=& \be P_{\text{M}}^{+} - \g P_{\text{S}}^{+} -\a_S P_{\text{S}}^{+}  + \a_S P_{\text{S}}^{-}  \, , \nn\\
\p_t P_{\text{S}}^{-} &=& \be P_{\text{M}}^{-} - \g P_{\text{S}}^{-}  - \a_S P_{\text{S}}^{-}  +\a_S P_{\text{S}}^{+}  \, .
\eea 
To obtain the effective diffusion coefficient $D_{\text{eff}}$ of these particles, one could exploit the fast time scales 
present in Eqs.~(\ref{eq:main1d}) using the method of adiabatic elimination detailed in~\cite{nava2018markovian}.   
Here, instead, we make use of an alternative, exact method that deals with all time scales involved.
Since $\la \mathbf{x} \ra(t) = \int_0^t dt'  \la \mathbf{u}(t') \ra$ and 
$\la \mathbf{x}^2 \ra(t) = \int_0^t dt'  \int_0^t dt'' \la \mathbf{u}(t') \cdot \mathbf{u}(t'') \ra$, 
we compute $D_{\text{eff}}$ as $D_{\text{eff}} = \lim_{t\to\infty} \frac{\la \mathbf{x}^2 \ra}{2d t}$, where $d$ is the dimension of the space. 
The challenge is then to calculate $\la \mathbf{u}(t) \cdot \mathbf{u}(t')\ra$. 
This quantity can be computed as follows: 
\bea
\label{eq:corr}
\la \mathbf{u}(t) \cdot \mathbf{u}(t')\ra &=&
\nonumber \sum_{q_1,q_2}\sum_{\s_1,\s_2}\!\!  v[q_1]v[q_2] \, {\sv}_1 \cdot {\sv}_2\, \\
&\times& P(\sv_2, q_2, t | \sv_1, q_1, t') P^{\sv_1*}_{q_1}\, ,
\eea
where $P(\sv_2, q_2, t | \sv_1, q_1, t')$ for $t\geq t'$ and $P^{\sv_1*}_{q_1}$ are, respectively, a conditional and a steady state pdf. 
Note that neither $P(\sv_2, q_2, t | \sv_1, q_1, t')$ nor $P^{\sv_1*}_{q_1}$ depend on $\xv$, and are the solutions of  
\bea
\label{eq:noSpace}
\partial_t \mathbf{P} =A\, \mathbf{P} \,
\eea
where $\mathbf{P} \equiv (P_{\text{M}}^{+}, P_{\text{M}}^{-}, P_{\text{S}}^{+}, P_{\text{S}}^{-})$, with 
\bea
A  = \begin{pmatrix}
-\alpha_M - \beta & \alpha_M & \gamma & 0\\
\alpha_M & -\alpha_M - \beta & 0 & \gamma \\
\beta & 0 & -\alpha_S-\gamma & \alpha_S \\
0 & \beta & \alpha_S & -\alpha_S - \gamma 
\end{pmatrix} 
\eea
Note that Eq.~(\ref{eq:noSpace}) is nothing else than Eq.~(\ref{eq:main1d}) without the terms involving spatial derivatives.  
From Eq.~(\ref{eq:noSpace}), $P(\sv_2, q_2, t | \sv_1, q_1, t')$ for $t\geq t'$ is obtained using  
$$\mathbf{P}(t')=(\la \delta_{+,\sv_1}\delta_{M,q_1} \ra, \la \delta_{-,\sv_1}\delta_{M,q_1} \ra, \la \delta_{+,\sv_1}\delta_{S,q_1} \ra,  \la \delta_{-,\sv_1}\delta_{S,q_1} \ra) $$ 
with $\d_{ij}$ denoting the Kronecker delta function, while $P^{\sv_1*}_{q_1}$ is the steady state of Eq.~(\ref{eq:noSpace}). 
We observe that in Eq.~(\ref{eq:corr}), the only non-zero terms correspond to $q_2=q_1=M$~\footnote{The Supplementary Information illustrates the Taylor-Kubo calculation of effective diffusivity using a simple 1D run-and-tumble model, compares diffusivities across different S-\&-G strategies, and provides a schematic overview of the models.
}.

In the following, we analyze the 4 models mentioned above. 
In model 1, we consider a simple dynamics in which the particle, when actively moving, stochastically switches from $\sv=+1$
to  $\sv=-1$ and vice-versa until it eventually stops.  After a stop, the particle moves with probability $1/2$ in the direction 
$\sv=+1$ and with $1/2$ in direction $\sv=-1$. This implies that the particle has no recollection of its moving direction $\sv$ previous to stop: the transition from $M$ to $S$ results in the resetting of $\sv$. 
Mathematically, it is equivalent to assuming that $P_{\text{S}}^{+}=P_{\text{S}}^{-}=P_{\text{S}}$, and thus we can describe the dynamics using only 3 fields, and redefine $\mathbf{P} = (P_{\text{M}}^{+}, P_{\text{M}}^{-}, P_{\text{S}})$ 
and reduce $A$ to: 
\bea
\label{eq:Asmall}
A  = \begin{pmatrix}
-\alpha_M - \beta & \alpha_M & \gamma/2 \\
\alpha_M & -\alpha_M - \beta &  \gamma/2 \\
\beta & \beta &  -\gamma 
\end{pmatrix} 
\eea
After solving Eq.~(\ref{eq:noSpace}) using Eq.~(\ref{eq:Asmall}) and computing the velocity autocorrelation,  
Eq.~(\ref{eq:corr}),  we arrive at
\bea
\label{eq:corrM1}
\la \uv(t) \cdot \uv(0)\ra = 
\la v^2\ra e^{-(2\a+\be)t}\, ,
\eea
where $\a_M=\a$, $\la v^2\ra = v_M^2 [P^{+*}_M + P^{-*}_M]$ with the steady-state 
probabilities $P^{+*}_M= P^{-*}_M= P^{\ast}=\hf\,\f{\g}{\be+\g}$ that are common to the 4 models. Eq.~(\ref{eq:corrM1}) yields an effective diffusion coefficient:
\begin{align}
D_{\rm eff}^{(1)} = \f{\la v^2\ra}{2\a+\be}.
\end{align}

The other three cases, models 2 to  4,  cannot be reduced to fewer than 4 fields.  
In model 2, the  polarity $\sv$ switches with rate $\a$ only in the $M$ state, implying  $\a_S=0$ and $\a_M=\a>0$. 
Consequently, the transition from $M$ to $S$ does not erase the memory of the heading direction before the stop. 
Following the procedure depicted above, we obtain: 
\bea
\label{eq:corr1D2}
\la \uv(t) \cdot \uv(0)\ra =   
\f{\la v^2\ra}{2s} [ (s+q)e^{-\l_3 t} +(s-q)e^{-\l_4 t} ]
\eea
where  $q=2\a+\be-\g$, $r=2\a+\be+\g$, $s=\sqrt{r^2 - 8 \a\g}$, inverse correlation times $\l_3=\hf(r+s)$, $\l_4=\hf(r-s)$. 
Integrating Eq.~(\ref{eq:corr1D2}), and taking the long-time limit, we obtain  the effective diffusivity:
\bea
D_{\rm eff}^{(2)} = \f{\la v^2\ra}{2\a}. 
\eea

In model 3, on the other hand, $\sv$ switches with  rate $\a$ both in the $M$ and $S$ states, and thus  $\alpha_M=\alpha_S=\alpha$. This implies that if the average time the particle remains in the stop state, $\gamma^{-1}$, 
is larger than the characteristic flipping time $\alpha^{-1}$, the transition to stop tends to erase the memory of the heading direction prior to the stop. Otherwise, the particle is likely to remember the previous moving direction. 
For this model, the velocity autocorrelation reads: 
\bea
\label{eq:corr1Dm3}
\la \uv(t) \cdot \uv(0)\ra =  
 \f{\la v^2\ra}{\be + \g} [\g e^{-\l_2 t} + \be e^{-\l_4 t}]\, ,
\eea
where $\l_2=2\a$ and $\l_4=(2\a+\be+\g)$.  
From Eq.~(\ref{eq:corr1Dm3}), we arrive at the effective diffusion coefficient: 
\bea
D_{\rm eff}^{(3)} = \f{\la v \ra^2}{2\a} + \f{\la \d v^2\ra}{2\a+\be+\g}
\eea
where $\la v \ra= v_M 2P^\ast 
= v_M \, \g/(\be+\g)$ and $\la \d v^2\ra =\la v^2\ra -\la v\ra^2 = v_M^2\, \be \g/(\be + \g)^2$.  
Note the increase in diffusivity due to speed fluctuations, cf.~\cite{peruani2007self, Shee2022}.

Finally, in model 4, the polarity $\sv$ remains fixed in the $M$ state ($\alpha_M=0$) and flips with rate $\alpha_S=\alpha$ only in the $S$ state. 
This implies that the memory of the heading direction can only be reset during a stop. This occurs with 
high probability when the average stop time $\gamma^{-1}$  is larger than  $\alpha^{-1}$. 
Solving Eq.~(\ref{eq:noSpace}) and replacing it into Eq.~({\ref{eq:corr}}), we get the velocity correlation:    
\bea
\la \uv(t) \cdot \uv(0)\ra =  \f{\la v^2\ra}{2s} [ (s\!+\!f)e^{-\l_3 t} \!+\!(s\!-\!f)e^{-\l_4 t} ]
\eea
where  $f=-2\a+\be-\g$, $r=2\a+\be+\g$, $s=\sqrt{r^2 - 8 \a\be}$, $\l_3=\hf(r+s)$,  and $\l_4=\hf(r-s)$. 
Thus, the effective diffusivity reads: 
\bea
D_{\rm eff}^{(4)} = \la v^2\ra \left( \f{2\a+\g}{2\a \be}\right)\,.
\eea
The models and the main results are summarized in Table-\ref{table}. 
Note that in~\cite{slowman2017exact} the exact solution of a run-and-tumble model in  1D with a finite tumble duration has also been reported, 
but where, in contrast to model 4, the probability of polarity inversion was assumed uncorrelated with the tumbling time. 

{\em Two-dimensional systems.--} 
In the following, we generalize the 4 S-\&-G models to 2D. 
For all the models, the temporal evolution of the particle position  is given as before by Eq.~(\ref{eq:xdot}), 
but where $\sv$ is not a discrete variable adopting two ${+1,-1}$, but a unit vector 
$\sv(\h)\equiv(\cos\h, \sin\h)$, whose orientation is encoded by the angle $\h$. 
On the other hand, the transition from the states $M$ and $S$ is governed as before by Eq.~(\ref{eq:behavior}). 
It is the dynamics of the now continuum polarity $\sv$ encoded by $\h$ that needs to be specified: 
\begin{align}
\label{eq:2Dspin}
\dot{\theta} = \sqrt{2\,D[q]\,}\eta(t)\, ,
\end{align}
where $\eta(t)$ is a white, $\delta$-correlated noise, and $D[q]$ is defined as 
$D[q=M]=D_M$ and $D[q=S]=D_S$, with $D_M$ and $D_S$ angular diffusion coefficients. 
Note that transitions $S\to M$ may also involve $\theta(t)$ updates. 
This occurs, for instance, in model 1.  
The transition $S \to M$ involves a resetting of $\sv$: the particle picks up a random direction $\sv$, 
choosing $\theta$ with equal probability from $[0,2\pi)$ when resuming active motion, and thus erasing the memory of the heading direction prior to the stop. 

In the following, we look for a description of the dynamics in terms of the 
probability distribution functions  of $M$ and $S$ states: $P_M(\theta,t)$ and $P_S(\theta,t)$, respectively. 
For model 1, where for simplicity we assume $D_S=0$, the corresponding Fokker-Planck equations read: 
\bea
\label{eq:model1_2D}
\p_t P_M &=& D_M \p_\h^2 P_M - \be P_M + \frac{\g}{2\pi} \int\,d\theta' \,P_S(\theta',t)\nn\\
\p_t P_S &=&  \be P_M - \g P_S.
\eea
Using that  $\int d\theta' \,[P_S+P_M] = 1$,  Eq.~(\ref{eq:model1_2D}) 
can be recast as:
\bea
\label{eq:model1_2D_SIMP}
\p_t P_M \!=\! D_M \p_\h^2 P_M \!-\! \be P_M \!+\! \frac{\g}{2\pi} \left[1\!-\!\!\int d\theta' P_M(\theta')\right]
\eea
Inserting into Eq.~(\ref{eq:model1_2D_SIMP}) the ansatz  $P_M\!=\!\frac{a_0(t)}{2} \!+\! \sum_{n=1}^{N} \left[a_n(t) \cos(n\theta)\!+\! b_n(t) \sin(n\theta)\right]$, we obtain equations and expressions for $a_n$ and $b_n$. 
Solving for $P_M(\theta, t)$, we compute the velocity autocorrelation for model 1 using Eq.~(\ref{eq:corr}): 
\bea
\la \uv(t)\cdot\uv(0)\ra= {\la v^2\ra} e^{-(D_M + \beta)t}\, ,
\eea
where $\la v^2\ra = v_M^2 P_M^\ast$ with $P_M^\ast = \g/(\be+\g)$, 
and its associated diffusivity:  
\bea
D_{\rm eff}^{(1)} = \frac{\langle  v^2\rangle}{2 D_M}  \frac{D_M}{D_M+\beta}\, . 
\eea
Note that this expression can be generalized for chiral particles with S-\&-G dynamics -- a situation of prime relevance to understand 
bacterial surface exploration~\cite{perez2019bacteria} -- by adding a term $-\w \p_\h P_M$ in Eq.\eqref{eq:model1_2D_SIMP}. 
This introduces oscillations in the velocity autocorrelation
\bea
\la \uv(t)\cdot\uv(0)\ra= {\la v^2\ra} e^{-(D_M + \beta)t} \cos (\w t)\, ,
\eea
and reduces the effective diffusivity to
\bea
D_{\rm eff}^{(1)} = \frac{1}{2}  \frac{D_M+\beta}{(D_M+\beta)^2+\w^2} \langle  v^2\rangle\, .
\eea

For the remaining three models, the Fokker-Planck equations take the form:  
\bea
\label{eq:modelS_2D}
\p_t P_M &=& D_M \p_\h^2 P_M - \be P_M + \g P_S\nn\\
\p_t P_S &=& D_S \p_\h^2 P_S + \be P_M - \g P_S.
\eea
Eq.~(\ref{eq:modelS_2D}) can be solved on a Fourier basis. One obtains the correlation function,
\begin{align}
\label{eq:corr_2d}
&\la \uv(t)\cdot\uv(0)\ra= \langle v[q(t)]v[q(0)] \cos(\theta(t)-\theta(0)) \rangle\nn\\ 
&= \f{\la v^2\ra}{\l_+^{(1)} - \l_-^{(1)}}
[(\l_+^{(1)}-d_1)e^{\l_+^{(1)} t} - (\l_-^{(1)}-d_1)e^{\l_-^{(1)} t}]
\end{align}
where $\la v^2\ra = v_M^2 P_M^\ast$ with $P_M^\ast= \g/(\be+\g)$, $d_1= -(D_S+\g)$, and eigenvalues corresponding to $n=1$ mode:  
\bea
\nonumber
\l_{\pm}^{(1)} &=& -\hf(D_M+D_S+\be+\g) \nn\\
&&\pm \hf[(D_M-D_S+\be-\g)^2+4\be \g]^\hf \nn\, .
\eea 

The effective diffusion constant, obtained from Eq.~(\ref{eq:corr_2d}), takes the form: 
\bea
\label{eq:genericD}
D_{\rm eff} = \f{\la v^2\ra}{2} \f{-d_1}{\l_+^{(1)} \l_-^{(1)}}. 
\eea
We apply Eq.~(\ref{eq:genericD}) to models 2, 3, and 4. 
In model 2, the polarity $\sv$ only evolves in the $M$ states, 
implying that $D_M= D_r$ and $D_S=0$, and thus: 
\bea
D_{\rm eff}^{(2)} = \f{\la v^2\ra}{2 D_r}. 
\eea
In model 3, the dynamics of $\sv$ is identical in both states, $M$ and $S$, 
i.e., $D_M=D_S=D_r$, and in consequence: 
\bea
D_{\rm eff}^{(3)} = \f{\la v \ra^2}{2 D_r} + \hf \f{\la \d v^2\ra}{D_r+\be+\g} \, , 
\eea 
where speed fluctuations $\la \d v^2\ra = \la v^2 \ra - \la v \ra^2  = v_M^2 P_M^\ast P_S^\ast$ with $P_S^\ast= \be/(\be+\g)$.  
Finally, in model 4, the memory of the heading direction $\sv$ can 
be lost only in the $S$ state since $D_M=0$ and $D_S=D_r$. 
This leads to a diffusivity: 
\bea
D_{\rm eff}^{(4)} = \f{\la v^2\ra}{2 D_r} \f{D_r+\g}{ \be}.
\eea

\begin{figure}
\begin{center}
\resizebox{\columnwidth}{!} {\includegraphics{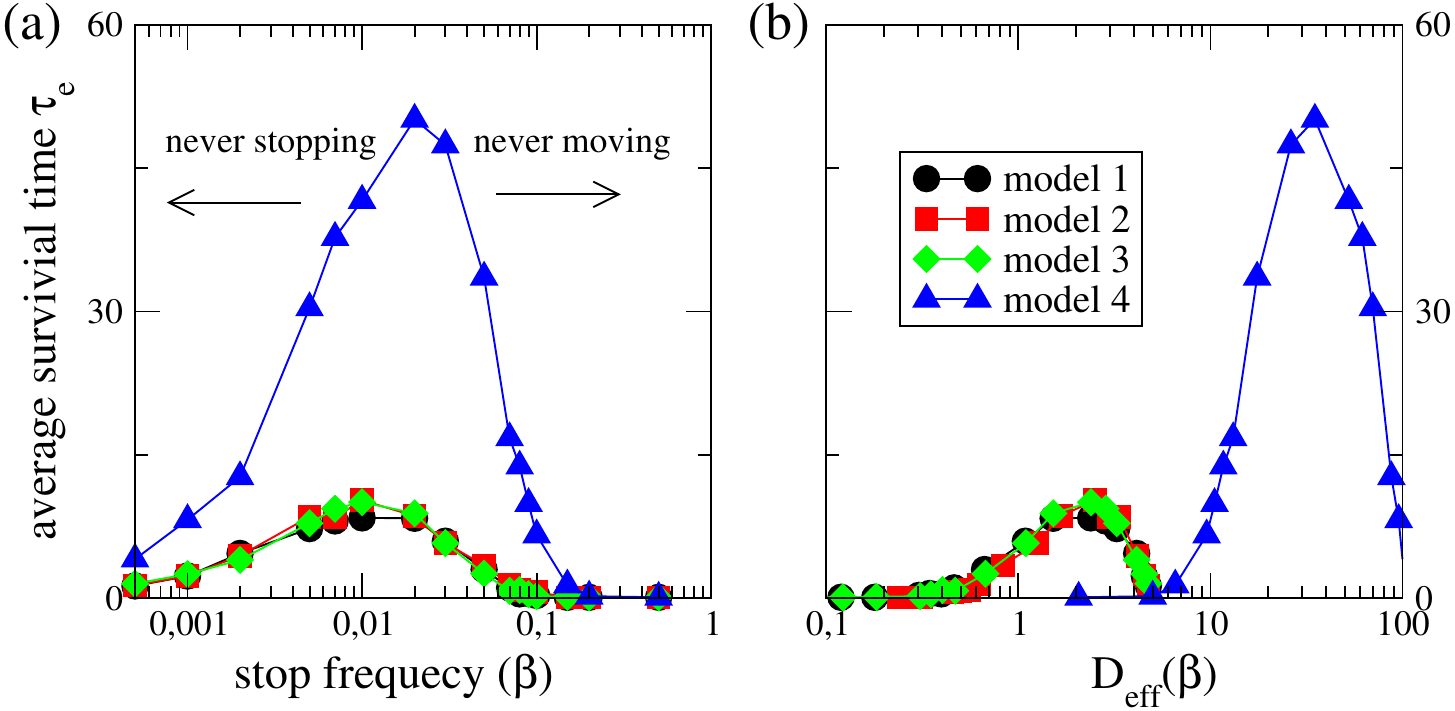} } 
\caption{
(Color online) Average survival time $\tau_e$ vs stop frequency $\beta$ in (a), and vs $D_{\rm eff}$ in (b), for the 4 models in 1D. 
Each point is an average over $10^4$ realizations. 
Parameters:  for motility $v_M=1$, $\gamma=0.01$, $\alpha_M=\alpha_S=0.1$, and for metabolic dynamics and food distribution $\mu = 0.005$, $\kappa_f = 1$, $e_0=2$, $\ell_f = 125$, and $R_f=5$ (see text).  
 }
\label{fig:1}
\end{center}
\end{figure}
{\em Survival time.--} 
In most biological systems, from bacteria to sheep, these two (behavioral) states, i.e., $M$ and $S$, are associated with very different behavioral needs. 
For instance, the stop state ($S$) may correspond to resting or probing the environment for nutrients (e.g., grass quality in sheep~\cite{gomez2022intermittent}) or suitable colonization sites (e.g., in bacteria~\cite{perez2019bacteria}).
It is only by incorporating the ecological dimension to the problem that the relevance of S-\&-G dynamics becomes undeniably evident. 
To this end, we extend our model by introducing minimal, generic metabolic dynamics: the active walker expends energy at a constant metabolic rate $\mu$, and during the stop state ($S$), it can uptake energy if nutrients are encountered at position ${\bf x}$. We assume a spatial distribution of discrete energy units $e_0$, spaced by a typical distance $\ell_f$. When the walker is in state $S$ at close proximity of a nutrient unit, i.e., within a distance $R_f$, it consumes it with a probability rate $\kappa_f$, removing it from the environment. The energy intake is described by $\varepsilon_{in}(t) = \sum_i e_0 \delta(t - t_i)$, where $t_i$ are the instances of nutrient uptake. The internal energy dynamics is thus given by $\dot{E} = -\mu + \varepsilon_{in}(t)$.
The active walker is considered alive as long as $E(t)>0$.  At  $t=0$, the walker is located between two nutrient sources with energy $E_0=1$. 
The total average survival time is $\tau = \tau_e + E_0/\mu$, where $E_0/\mu$ is time the walker can live with the initial energy $E_0$ 
and $\tau_e$ is the average time it survives beyond that initial period $E_0/\mu$. 

Figure~\ref{fig:1} reveals that there exists an optimal stopping frequency $\beta$ -- and thus an optimal effective diffusion coefficient $D_{\text{eff}}$ -- that maximizes the average survival time. This maximization reflects a trade-off between nutrient uptake (favored by stopping) and exploration (enhanced by motion). While $D_{\text{eff}}$ decreases monotonically with increasing $\beta$ and is maximized as $\beta \to 0$ (i.e., when the walker never stops), survival time vanishes in this limit due to the lack of energy uptake. Hence, maximizing $D_{\text{eff}}$ does not equate to maximizing survival. Nevertheless, for strategies with identical mean run and stop durations (fixed $\beta$ and $\gamma$), higher $D_{\text{eff}}$ leads to longer survival, emphasizing the critical role of microscopic dynamics in shaping optimal survival strategies.
Note that one obtains similar results if the energy units $e_0$ are allowed to regrow or if the metabolic expenditure is assumed proportional to $E$, i.e., $- \mu E$. 
Arguably, the optimal behavior of survival time with the stop frequency and diffusivity is generic and ubiquitous in nature: not moving leads to fast depletion of local nutrients and thus to death, while constant motion does not provide the necessary time to the organism for nutrient uptake and yields the same outcome, i.e., death.  
In short, the obtained results prove that  there exist (i) an optimal stopping frequency and (ii) 
a non-trivial value of the diffusion coefficient $D_{\text{eff}}$, which itself depends on $\beta$, 
that maximize survival.

Although the survival time problem studied here may appear related to intermittent search strategies~\cite{benichou2011intermittent}, it is fundamentally different in scope and mechanism.
A more relevant comparison is with the starving random walk model~\cite{Benichou2018, Regnier2024}, yet even this analogy has important limitations. In the starving random walk, the walker's energy is reset to a fixed value upon each food encounter. In contrast, our model allows energy to accumulate indefinitely, without an upper limit. Here, survival time corresponds to the first passage time of the internal energy \(E\), which performs a random walk restricted to the positive half-line---unlike the bounded dynamics in~\cite{Benichou2018}. Crucially, maximizing survival time in this context is not equivalent to minimizing the first-passage time to a target.

It is worth stressing that although we assume Poissonian transitions between states, our framework can be readily extended to incorporate more complex switching dynamics, such as transitions via multiple internal states~\cite{perez2019bacteria, Alirezaeizanjani2020Bacteria, Gupta2019, nava2018markovian}, renewal processes~\cite{kurzthaler2024characterization}, active speed fluctuations~\cite{Schienbein1993, peruani2007self, Shee2022}, or inertial relaxation~\cite{Patel2023, Devereux2021, Rabault2019}.

{\em Acknowledgments} - 
F.P. acknowledges financial support from C.Y. Initiative of Excellence (grant Investissements d'Avenir ANR-16-IDEX-0008), INEX 2021 Ambition Project CollInt and Labex MME-DII, projects 2021-258 and 2021-297. { FP thanks Robert Grossmann for the insightful discussion on adiabatic elimination and useful comments.}  D.C. acknowledges C.Y. Cergy Paris Universit{\'e} for a Visiting Professorship, SERB (India) for grant MTR/2019/000750, DAE (India) for grant 1603/2/2020/IoP/R\&D-II/150288, and ICTS-TIFR, Bangalore, for an Associateship.

\bibliographystyle{apsrev4-2}

%


\section*{Supplemental Material: Illustration of the Taylor-Kubo calculation using the 1D RTP model}

 \begin{figure*}[t]
\begin{center}
\includegraphics[width=0.9 \linewidth]{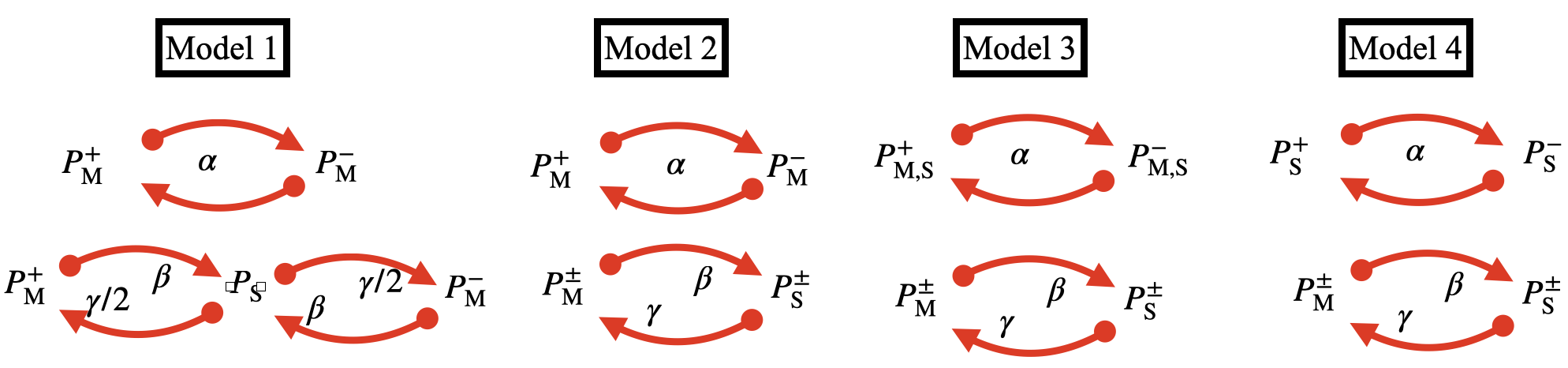}
\caption{Schematics of the four models in terms of transition probabilities between various states as described in the main text.}
\label{fig_cartoon}
\end{center}
\end{figure*}

Here we illustrate the method of calculating the velocity autocorrelation and diffusivity that we used in the main text. We do this by using the example of a simple RTP with right and left moving states $P_\pm (x,t)$ evolving as
\bea
\p_t P_+ &=& -v_0 \p_x P_ + -\a P_+ + \a P_-   \nn\\
\p_t P_- &=& v_0 \p_x P_- - \a P_- + \a P_+ .  
\label{eq_2state}
\eea 
Note that the $\pm$ states are independent of the speed $v_0$. We utilize this fact to write down the general solution $(P_+, P_-)^T (t) = a_1 e^{\l_1 t} |\psi_1\rangle + a_2 e^{\l_2 t} |\psi_2\rangle$ in terms of the eigenvalues $\l_{1,2}$ and eigenvectors $|\psi_{1,2}\rangle$ corresponding to the transition matrix $( (-\a,\a)^T, (\a,-\a)^T )$. The constants $a_{1,2}$ are determined using the initial condition, i.e.  $P_+(x,0)$ and $P_-(x,0)$. As a result, we obtain the four transition probabilities $P(+,t | \pm, 0)$, $P(-,t | \pm, 0)$ and the two steady-state solutions $P^{(s)}_\pm=1/2$  in the limit of $t \to \infty$. Thus, we can calculate the steady-state auto-correlation function $\la v(t) v(0) \ra = v_0^2 [ P(+,t | +, 0) + P(-,t | -, 0) - P(-,t | +, 0) -  P(+,t | -, 0)] P_\pm^{(s)}$, which leads to the mean-squared displacement $\la x^2(t)\ra = \int_0^t dt_1 \int_0^t dt_2 \la v(t_1) v(t_2)\ra$. Using explicit calculations, one can easily check that  
$\la v(t_1) v(t_2) \ra = v_0^2 \exp(-2\a |t_2-t_1|)$, and 
$\la x^2(t)\ra 
= \f{v_0^2}{2\a^2} (e^{-2 \a t}-1+2\a t)$. This predicts a ballistic-diffusive crossover at $t=1/2\a$ and an asymptotic effective diffusion constant $D_{\rm  eff} = \lim_{t\to\infty} \la x^2(t)\ra/2t = v_0^2/2\a$.

\section{Comparison of models 1 to 4}
We studied 4 S-\&-G models that, though differ only in subtle details of their dynamics,  
exhibit distinct large-scale transport properties, i.e., diffusivity. 
These models correspond to different experimental situations. 
The resetting of the heading direction associated to a stop in model 1 is found in bacterial surface exploration and Quincke rollers under an AC field~\cite{perez2019bacteria,karani2019tuning, pradillo2019quincke}. 
Active motion with pauses that keep the memory of the active moving direction, i.e. model 2, occurs in several animal systems, where stops are often used to scan potential predators~\cite{kramer2001behavioral, gomez2022intermittent}. 
In cell migration, changes in the moving direction and cell polarization occur when the cell moves and stops, as in model 3~\cite{Schienbein1993, Vaidziulyte2022}. 
Finally, model 4 describes straight active runs follow by reorientation events proportional to the stop duration, which is a characteristic feature 
of bacterial run-and-tumble motion~\cite{figueroa20203d, Wadhwa2022bacteria}. 
In the following, we compare the S-\&-G strategy against the classical dynamics of active particles that move at constant speed $v_0$ and change the heading direction at rate $\a$, displaying diffusivity $D_{\rm eff}^{(0)}=v_0^2/2\a$.

We start by ensuring that both S-\&-G and persistent active particles exhibit the same mean speed $\la v \ra=v_M \f{\g}{\be+\g}=v_0$. 
Under this condition, the diffusivity ratios are  
${D_{\rm eff}^{(1)}}/{D_{\rm eff}^{(0)}}= \frac{\gamma}{\gamma+\beta}[\frac{2\a}{2\a+\be}]$, ${D_{\rm eff}^{(2)}}/{D_{\rm eff}^{(0)}}= \frac{\gamma}{\gamma+\beta}$, 
$\f{D_{\rm eff}^{(3)}}{D_{\rm eff}^{(0)}}= \left[ 1 +\f{\be}{\g}\f{2\a}{2\a+\be+\g} \right]$, and 
${D_{\rm eff}^{(4)}}/{D_{\rm eff}^{(0)}}= \frac{\gamma}{\gamma+\beta} \frac{2\a+\g}{\be}$, 
where $D_{\rm eff}^{(i)}$ is the diffusivity of model $i$.  
This implies that if we compare systems with the same average speed, the diffusivity of S-\&-G particles can be larger than that of persistently active particles, but the energy dissipation rate of S-\&-G particles is also larger.  
Let us recall that assuming a viscous drag, the dissipation rate $q$ of an active particle moving at speed $v$  is $q \propto \langle v^2 \rangle$. Thus, the comparison of dissipation rates  can be expressed, for all models, as 
$q^{i}/q_0:= \la v^2\ra/v_0^2 = 1+\be/\g > 1$ and $\la \d v^2\ra/v_0^2 = \be/\g$, implying 
that the excess of dissipation can be attributed to the velocity fluctuations of S-\&-G particles. 

Now, instead of comparing systems with mean speed, we analyze systems that display the same dissipation rate by 
requesting that $\la v^2\ra=v_M^2 \g/(\be +\g)=v_0^2$. 
Under this constraint, we find that ${D_{\rm eff}^{(1)}}/{D_{\rm eff}^{(0)}}= [{2\a}/({2\a+\be})]<1$, 
 ${D_{\rm eff}^{(2)}}/{D_{\rm eff}^{(0)}}= 1$,  
$\f{D_{\rm eff}^{(3)}}{D_{\rm eff}^{(0)}}= \f{\g}{\be+\g}\left[ 1 +\f{\be}{\g}\f{2\a}{2\a+\be+\g} \right]<1$, 
and ${D_{\rm eff}^{(4)}}/{D_{\rm eff}^{(0)}}=\left[({2\a+\g})/{\be}\right]$. 
Note that assuming equal energy expenditure, i.e., dissipation power, 
 only model 4 -- which corresponds to the bacterial run-and-tumble motion -- can exhibit a larger diffusivity than persistently active particles. This suggests that by performing the run-and-tumble motion, bacteria achieve longer spatial exploration at the same power cost.

\end{document}